\documentclass{kapproc} 

\usepackage{t1enc}
\usepackage{procps} 
\usepackage[dvips]{graphicx}
\upperandlowercase
\kluwerbib 

\def\mbox{\hbox}           

\def\deg{\ifmmode ^\circ                
         \else $^\circ$
         \fi
         \hskip -0.1truecm}
\def\degd#1.#2{                         
               \ifmmode {#1^{\hskip 0.05em\circ}\hskip-0.42em.\hskip0.08em#2}
               \else {#1$^{\hskip 0.05em\circ}\hskip-0.42em.\hskip0.08em$#2}
               \fi
              }
\def\mind#1.#2{                         
               \ifmmode {#1^{\hskip 0.05em\prime}\hskip-0.35em.\hskip0.05em#2}
               \else {#1$^{\hskip 0.05em\prime}\hskip-0.35em.\hskip0.05em$#2}
               \fi
              }
\def\secd#1.#2{                         
               \ifmmode {#1^{\prime\prime}\hskip-0.46em.\hskip0.12em#2}
               \else {#1$^{\prime\prime}\hskip-0.46em.\hskip0.12em$#2}
               \fi
              }
\def\timsecd#1.#2{                      
                  \ifmmode {#1^{\rm s}\hskip-0.39em.\hskip0.08em#2}
                  \else {$#1^{\rm s}\hskip-0.39em.\hskip0.08em#2$}
                  \fi
                 }
\def\hms#1h#2m#3s{                      
                  \relax
                  \ifmmode #1^{\rm h}\,#2^{\rm m}\,#3^{\rm s}
                  \else \hbox{$#1^{\rm h}\,#2^{\rm m}\,#3^{\rm s}$}
                  \fi
                 }
\def\dms#1d#2m#3s{                      
                  \relax
                  \ifmmode #1^\circ\,#2^{\prime}\,#3^{\prime\prime}
                  \else \hbox{$#1^\circ\,#2^{\prime}\,#3^{\prime\prime}$}
                  \fi
                 }
\def\dmsd#1d#2m#3.#4s{                  
                      \relax
                      \ifmmode #1^\circ\,#2^{\prime}\,#3^{\prime\prime}
                               \hskip-0.46em.\hskip0.12em#4
                      \else \hbox{$#1^\circ\,#2^{\prime}\,#3^{\prime\prime}
                            \hskip-0.46em.\hskip0.12em#4$}
                      \fi
                     }
\def\hm#1h#2m{                          
              \relax
              \ifmmode #1^{rm h}\,#2^{\rm m}
              \else \hbox{$#1^{\rm h}\,#2^{\rm m}$}
              \fi
             }
\def\dm#1d#2m{                          
              \relax
              \ifmmode #1^\circ\,#2^{\prime}
              \else \hbox{$#1^\circ\,#2^{\prime}$}
              \fi
             }
\def\hmsd#1h#2m#3.#4s{                  
                      \relax
                      \ifmmode #1^{\rm h}\,#2^{\rm m}\,#3^{\rm s}
                               \hskip-0.39em.\hskip0.08em#4
                      \else \hbox{$#1^{\rm h}\,#2^{\rm m}\,#3^{\rm s}
                            \hskip-0.39em.\hskip0.08em#4$}
                      \fi
                     }
\def\hmd#1h#2.#3m{                  
                  \relax
                  \ifmmode #1^{\rm h}\,#2^{\rm m}
                           \hskip-0.55em.\hskip0.22em#3
                  \else \hbox{$#1^{\rm h}\,#2^{\rm m}
                        \hskip-0.55em.\hskip0.22em#3$}
                  \fi
                 }
\def\mg{\relax                          
        \ifmmode ^{\rm m}
        \else $^{\rm m}$
        \fi
       }
\def\mgd#1.#2{                          
              \relax
              \ifmmode #1^{\rm m}
                       \hskip-0.55em.\hskip0.22em#2
              \else \hbox{#1$^{\rm m}
                    \hskip-0.55em.\hskip0.22em$#2}
              \fi
             }

\def\la{\mathrel{\hbox{\rlap{\hbox{\lower4pt\hbox{$\sim$}}}\hbox{$<$}}}}
\def\ga{\mathrel{\hbox{\rlap{\hbox{\lower4pt\hbox{$\sim$}}}\hbox{$>$}}}}

\def\unitspace{\;}                      

\def\un#1{\ifmmode \unitspace\mbox{\rm #1} 
          \else $\unitspace$#1
          \fi}
\def\pun#1#2{\ifmmode \unitspace\mbox{\rm #1}^{#2} 
             \else $\unitspace$#1$^{#2}$
             \fi}

\def\Lsun{\ifmmode \un{L}_{\odot}     
          \else $\un{L}_{\odot}$
          \fi}
\def\Msun{\ifmmode \un{M}_{\odot}     
          \else $\un{M}_{\odot}$
          \fi}
\def\mum{\ifmmode \unitspace\mu\mbox{\rm m} 
         \else $\unitspace\mu$m
         \fi}
\def\sqarcsec{\ifmmode \unitspace\Box''    
              \else $\unitspace\Box''$     
              \fi} 

\def\Bp{\relax                            
        \ifmmode B_{||}                   
        \else $B_{||}$
        \fi}
\def\Bt{\relax                            
        \ifmmode B\!_{\perp}              
        \else $B\!_{\perp}$               
        \fi}
\def\Gcr{\relax                           
         \ifmmode \Gamma\!_{\rm cr}       
         \else $\Gamma\!_{\rm cr}$
         \fi}
\def\ICII{\relax                          
          \ifmmode I_{[\CII]}             
          \else $I_{[\CII]}$
          \fi}
\def\LHtwo{\relax                                 
           \ifmmode L_{\mbox{\rm\scriptsize H}_2} 
           \else $L_{\mbox{\rm\scriptsize H}_2}$  
           \fi}
\def\LLya{\relax                          
          \ifmmode L_{{\rm Ly}\,\alpha}   
          \else $L_{{\rm Ly}\,\alpha}$
          \fi}
\def\MHtwo{\relax                                 
           \ifmmode M_{\mbox{\rm\scriptsize H}_2} 
           \else $M_{\mbox{\rm\scriptsize H}_2}$  
           \fi}
\def\MHtwodot{\relax                                       
              \ifmmode \dot{M}_{\mbox{\rm\scriptsize H}_2} 
              \else $\dot{M}_{\mbox{\rm\scriptsize H}_2}$  
              \fi}                                         
\def\Mstardot{\relax                      
              \ifmmode \dot{M}_{\ast}     
              \else $\dot{M}_{\ast}$      
              \fi}
\def\nHI{\relax                                      
         \ifmmode n_{\mbox{\scriptsize\rm H\,\sc I}} 
         \else $n_{\mbox{\scriptsize\rm H\,\sc I}}$
         \fi}
\def\nHtwo{\relax                                
           \ifmmode n_{{\mbox{\scriptsize H}}_2} 
           \else $n_{{\mbox{\scriptsize H}}_2}$  
           \fi}
\def\rhostardot{\relax                         
                \ifmmode \dot{\rho}_{\ast}     
                \else $\dot{\rho}_{\ast}$      
                \fi}
\def\rhoZdot{\relax                          
             \ifmmode \dot{\rho}_{\rm Z}     
             \else $\dot{\rho}_{\rm Z}$      
             \fi}

\def\sou#1#2{\relax                       
             \ifmmode {\rm #1}\,{\rm #2}  
             \else #1$\,$#2
             \fi}

\def\NGC#1{\sou{NGC}{#1}}                

\def\Arp#1{\sou{Arp}{#1}}                

\def\qu#1#2{\relax                          
            \ifmmode #1_{\rm #2}            
            \else $#1_{\rm #2}$
            \fi}

\def\mbox{\hbox}           

\def\CO#1{\ifnum#1=0                    
           \ifmmode \mbox{\rm CO}
           \else {\rm CO}
           \fi
          \else
           \ifnum#1<15
            \ifmmode ^{#1}\mbox{\rm CO}
            \else $^{#1}${\rm CO}
            \fi
           \else
            \ifmmode \mbox{\rm C}^{#1}\mbox{\rm O}
            \else {\rm C}$^{#1}${\rm O}
            \fi
           \fi
          \fi}

\def\COp{\ifmmode \mbox{\rm CO}^+           
         \else {\rm CO}$^+$                 
         \fi}

\def\CS#1{\ifnum#1=0                    
           \ifmmode \mbox{\rm CS}
           \else {\rm CS}
           \fi
          \else
           \ifnum#1<15
            \ifmmode ^{#1}\mbox{\rm CS}
            \else $^{#1}${\rm CS}
            \fi
           \else
            \ifmmode \mbox{\rm C}^{#1}\mbox{\rm S}
            \else {\rm C}$^{#1}${\rm S}
            \fi
           \fi
          \fi}

\def\HCOp{\ifmmode \mbox{\rm HCO}^+          
          \else {\rm HCO}$^+$                
          \fi}
\def\Hthreep{\ifmmode \mbox{\rm H}_3^+         
             \else {\rm H}$_3^+$               
             \fi}

\def\Htwo{\ifmmode \mbox{\rm H}_2              
          \else {\rm H}$_2$                    
          \fi}

\def\HtwoO{\ifmmode \mbox{\rm H}_2\mbox{\rm O} 
           \else {\rm H}$_2${\rm O}            
           \fi}

\def\ion#1#2{\ifmmode \mbox{{\rm #1}}\,\mbox{{\sc #2}} 
        \else {\rm #1}$\,${\sc #2}
        \fi}

\def\HII{\ion{H}{ii}}

\def\rec#1#2{\if#2a                            
              \ifmmode \mbox{{\rm #1}}\alpha   
              \else {\rm #1}$\alpha$
              \fi
             \fi
             \if#2b
              \ifmmode \mbox{{\rm #1}}\beta
              \else {\rm #1}$\beta$
              \fi
             \fi
             \if#2g
              \ifmmode \mbox{{\rm #1}}\gamma
              \else {\rm #1}$\gamma$
              \fi
             \fi}

\def\Brg{\rec{Br}{g}}                          

\newcommand{\figref}[1]{Fig.~\protect\ref{#1}}

\newcommand{\eqref}[1]{Eq.~$\left(\protect\ref{#1}\right)$}

\begin{document}

\articletitle{Dissecting starburst galaxies\\ 
with infrared observations}

\author{Paul P.~van der Werf and Leonie Snijders}
\affil{Leiden Observatory\\
P.O.~Box 9513\\
NL - 2300 RA Leiden\\
The Netherlands}
\email{pvdwerf@strw.leidenuniv.nl, snijders@strw.leidenuniv.nl}

\begin{abstract}
The infrared regime contains a number of unique diagnostic features
for probing the astrophysics of starburst galaxies. After a brief
summary of the most important tracers, we focus in detail on the use of
emission lines to probe the upper part of the main sequence in a young
superstarcluster in the Antennae, and the compact, dusty starburst in
the nucleus of the nearby ultraluminous infrared galaxy $\Arp{220}$.
\end{abstract}

\section{Introduction}
Since stars form in the cores of dusty molecular clouds, it is not
surprising that optical obscuration forms a major stumbling block for
studying starburst galaxies. Since dust content and extinction correlate
with stellar luminosity (e.g., Kennicutt,
these proceedings), this is in particular true for the more
luminous starbursts. Thus, while optical and ultraviolet (UV) observations of
relatively unobscured regions in low to moderate luminosity starburst
galaxies still provide excellent astrophysical diagnostics (e.g.,
Leitherer, these proceedings), more luminous systems require observations
at near-infrared and longer wavelengths.

However, reduced extinction ($A_K\sim0.11A_V$) is not the only reason for
observing starburst galaxies in the infrared. The infrared regime also
contains an extensive set of diagnostic features which are ideal probes
of the parameters of the stellar population as well as its feedback on
the ambient gas:
\begin{enumerate}
\item
the Lyman continuum flux can be probed using hydrogen recombination
lines, principally from the Paschen and Brackett series (as well as by
thermal radio continuum emission);
\item
the temperature of the ionizing radiation field can be probed using
suitable ratios of helium and hydrogen recombination lines, as well as
by ratios of suitable combinations of fine-structure lines (e.g., 
\cite{Shields93}; \cite{Lumsdenetal03}; \cite{RigbyRieke04});
\item
a measure of the supernova rate can be obtained from the near-infrared
(near-IR) [$\ion{Fe}{ii}$] lines, as well as from non-thermal radio
continuum emission (e.g., \cite{VanDerWerfetal93}; \cite{AlonsoHerreroetal03});
\item
the age of the stellar population can be derived from any ratio of
tracers that probe different temporal phases of the starburst, such as
the \Brg equivalent width, effectively probing the relative importance
of O-stars and red supergiants (e.g., \cite{Leithereretal99});
\item
warm molecular gas, heated by shocks or by UV radiation, can be probed by
the $\Htwo$ near-IR rovibrational lines and the mid-infrared (mid-IR) 
rotational lines, as well as low-excitation fine-structure lines such as
the [$\ion{C}{ii}$] $158\mum$ line;
\item
extinction can be derived from ratios of hydrogen recombination lines,
from ratios of $\Htwo$ or [$\ion{Fe}{ii}$]
lines arising from the same upper level, and from analysis of ice and
silicate absorption features;
\item
a possible hidden active galactic nucleus (AGN) can be revealed by
high-excitation lines such as [$\ion{Si}{vi}$] $1.96\mum$;
\item
emission and absorption line 
kinematics can be used to derive dynamical masses, and to probe bulk
flows;
\item
finally, in luminous and ultraluminous infrared galaxies, the best
measure of the total luminosity of the starburst is provided by the
integrated far-infrared (far-IR) emission, which typically dominates the
bolometric energy output.
\end{enumerate}
However, care should be taken not to apply these diagnostics blindly,
since none of them is entirely straightforward, and some are only
usable in limited regions of parameter space. A complete discussion of
the caveats is beyond the scope of this paper, and the reader is
referred to the references above for more details. 

A detailed analysis, using a combination of these parameters, has been
done for the nearby starburst galaxy M82 
(\cite{ForsterSchreiberetal01}, \cite{ForsterSchreiberetal03}),
resulting in a detailed view of the spatial and temporal evolution of
this starburst. More limited studies using mostly near-IR data, have been
published of other nearby starbursts such as $\NGC{253}$
(\cite{Forbesetal93}, \cite{Engelbrachtetal98}), $\NGC{1808}$ 
(\cite{Kotilainenetal96}),
$\NGC{7552}$ (\cite{Schinnereretal97}) and $\sou{IC}{342}$ 
(\cite{Bokeretal97}).

In this paper we focus in particular on the use of near-IR hydrogen
recombination lines to derive the Lyman continuum flux of obscured
starbursts. We first discuss the spectral properties of a powerful young
superstarcluster in the Antennae ($\NGC{4038{/}4039}$) and then contrast
these to the spectral properties of the nearby ultraluminous infrared
galaxy (ULIG) $\Arp{220}$. The analysis is based on near-IR H- and
K-band spectra obtained with ISAAC at the ESO Very Large Telescope.

\begin{figure}[ht]
\hbox to \hsize{
\includegraphics[height=6cm,angle=90]{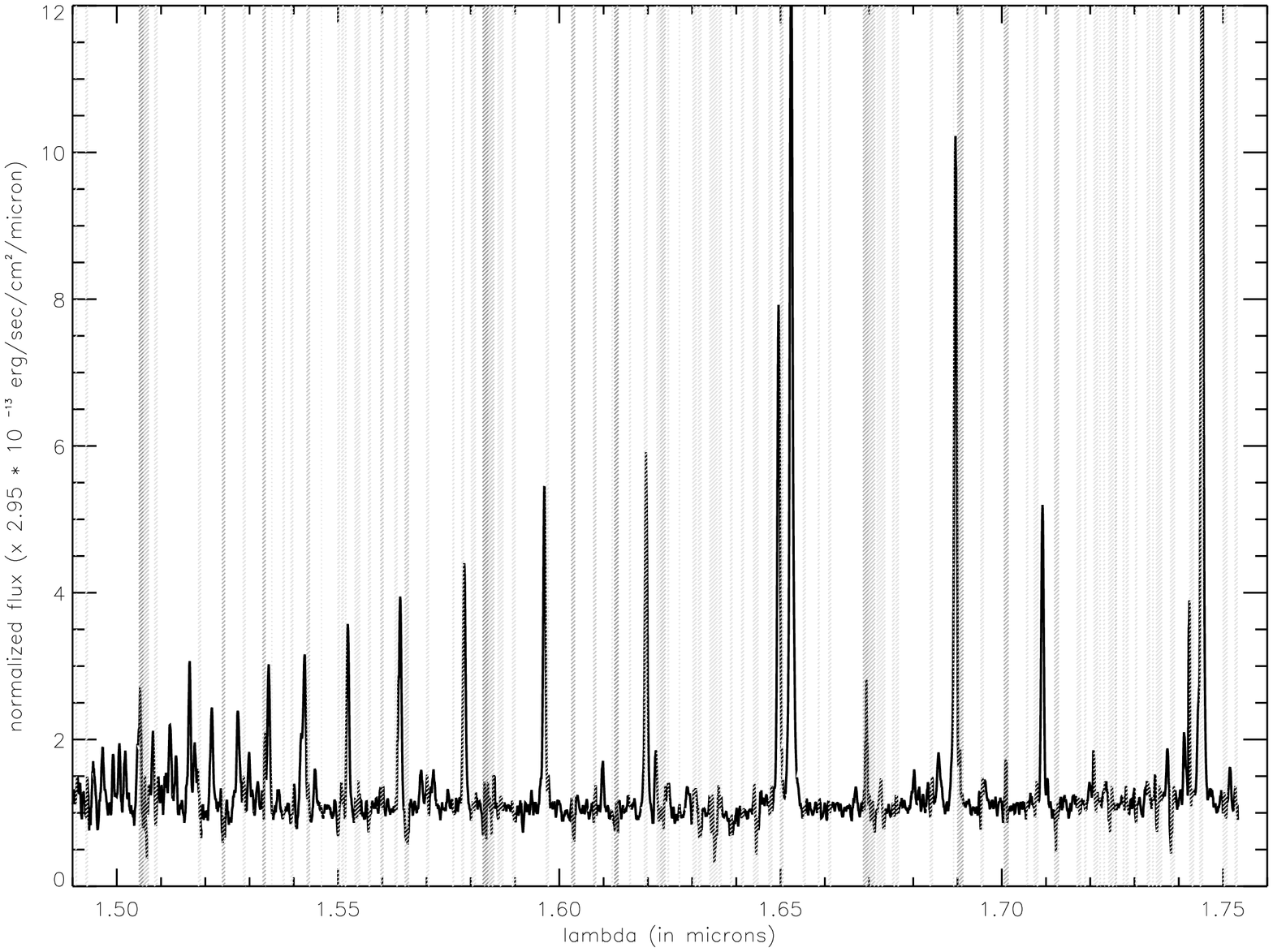}
\includegraphics[height=6cm,angle=90]{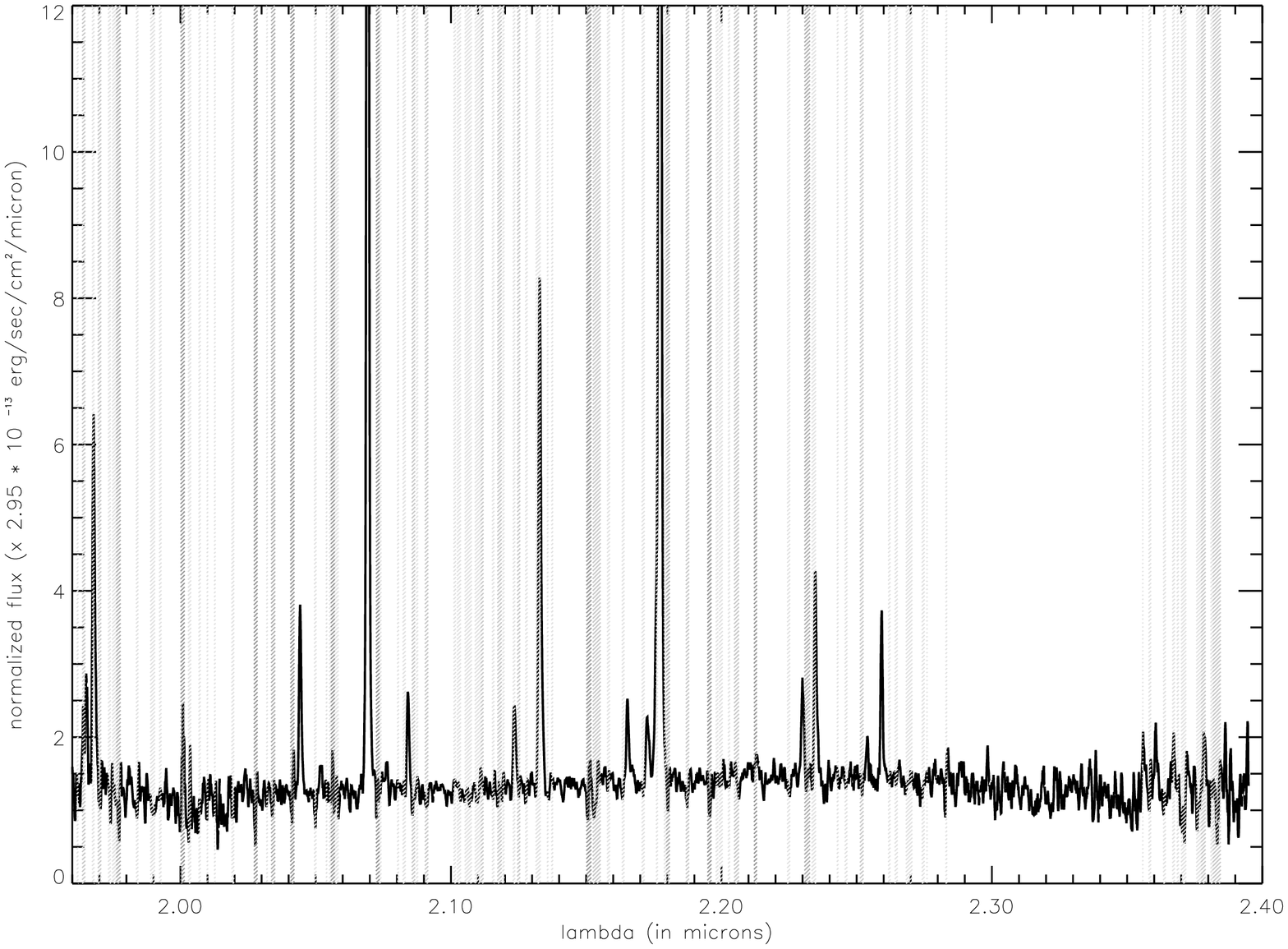}
}
\caption{H-band (left) and K-band (right) spectra of 
superstarcluster 80 in the Antennae. 
Shaded bands indicate spectral
regions affected by skylines.
\label{fig.Antspectra}
}
\end{figure}

\section{Near-IR spectra of a superstarcluster in the Antennae}

In \figref{fig.Antspectra} we present near-IR H- and K-band spectra of a
young, obscured superstarcluster in the Antennae. This cluster (object
number~80 in the notation of \cite{WhitmoreSchweizer95})
is a prominent mid-IR source,
producing about 15\% of the total $15\mum$ flux density of the entire
system (\cite{Mirabeletal98}).
The spectra show the typical features expected for obscured starforming
regions: the H-band is dominated by the Brackett series, with in
addition a bright [$\ion{Fe}{ii}$] line at $1.64\mum$, and a prominent
$\ion{He}{i}$ line at $1.70\mum$. The K-band is dominated by very strong
$\Brg$ and $\ion{He}{i}$ $2.06\mum$ emission (with weaker $\ion{He}{i}$ 
emission at
$2.11\mum$), and a number of $\Htwo$ rovibrational lines. 
CO absorption bands are visible at $2.32$ and $2.36\mum$, but are
extremely faint, as expected for this very young cluster
($\sim4\un{Myr}$, \cite{Gilbertetal00}).

The large number of hydrogen recombination lines detected allow an
accurate extinction determination. Fitting all lines simultaneously, we
obtain an extinction $A_K=0.5$, located in a foreground screen. This
geometry provides a significantly better fit than a model where
emitting and absorbing material are cospatial. After correction for
extinction, the derived Lyman continuum flux (under the usual
assumptions of case~B recombination in ionization bounded, dust-free
$\HII$ regions) is $Q_0=1.5\cdot10^{53}\pun{s}{-1}$, which implies the
presence of about 35000~O-stars within a volume with a half-light radius
of only $32\un{pc}$. The effective temperature
of the radiation field derived from the ratio of helium and hydrogen
recombination lines is $\qu{T}{eff}\sim38500\un{K}$, which corresponds to
the most massive stars having a Zero-Age Main Sequence 
spectral type of approximately O7.5. Given the derived cluster age,
stars with spectral types up to O4.5 could have been present, but in
that case $\qu{T}{eff}\sim47000\un{K}$ would be expected. This value is
closer to the $\qu{T}{eff}\sim44000\un{K}$ estimated from ratios of
mid-IR fine-structure lines (\cite{Kunzeetal96}). The disagreement with
the value derived from recombination lines is harmless, since the 
recombination line ratios are insensitive to $\qu{T}{eff}$ values above about
$40000\un{K}$. More detailed analysis
is required to see if this procedure can be used to establish the
presence of an upper mass cutoff on the initial mass function.

\begin{figure}[ht]
\hbox to \hsize{
\includegraphics[height=6cm,angle=90]{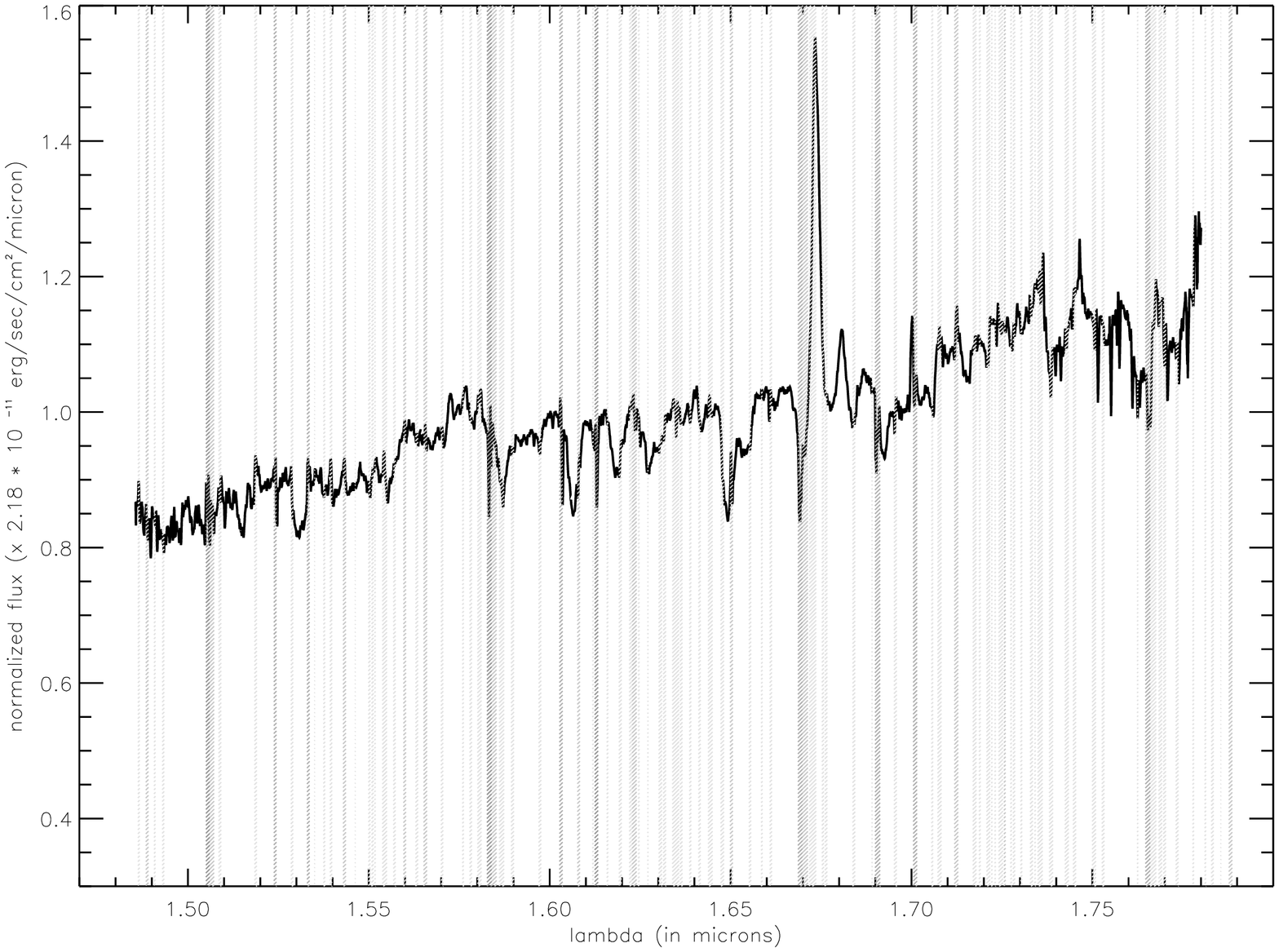}
\includegraphics[height=6cm,angle=90]{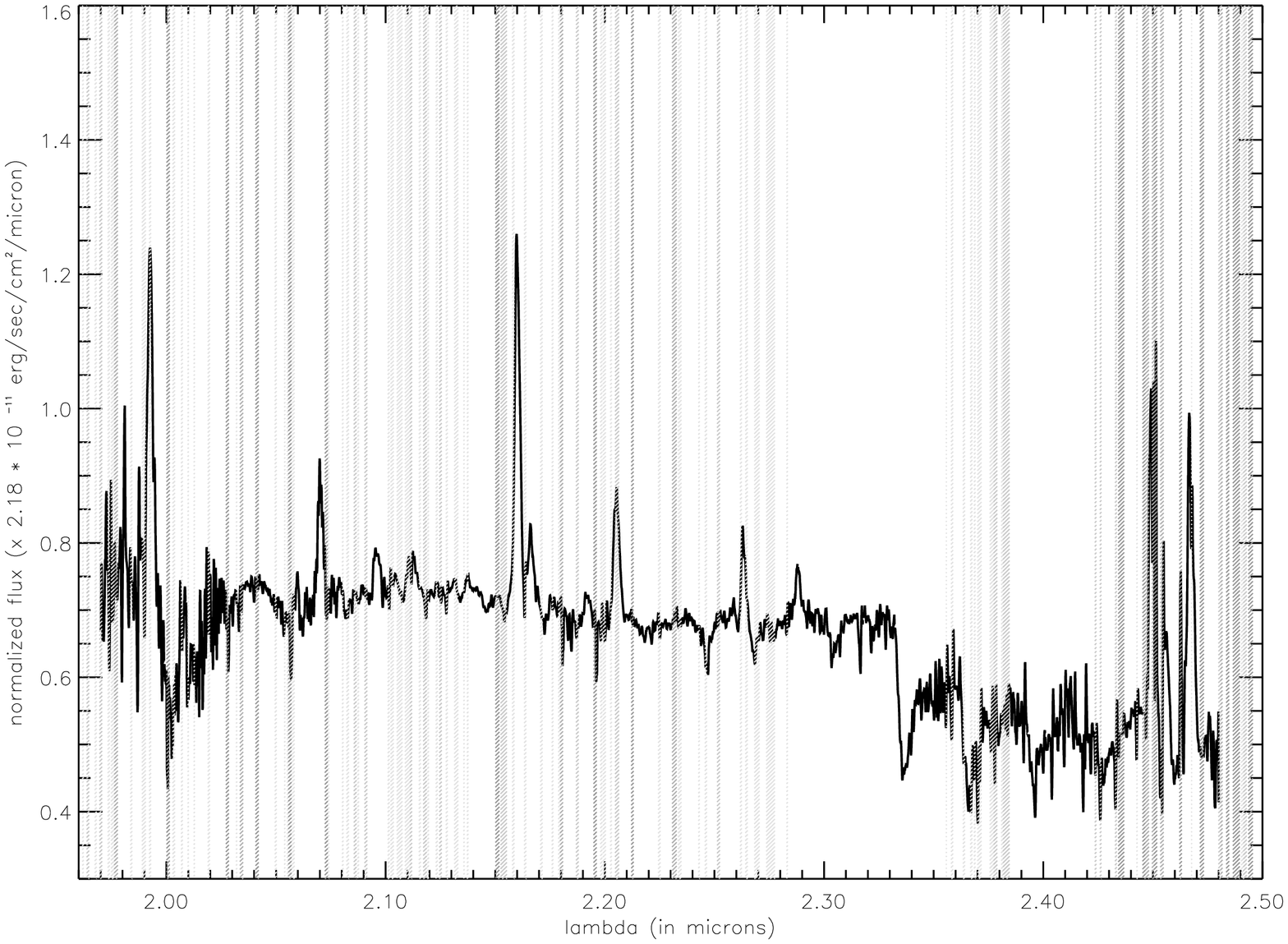}
}
\caption{H-band (left) and K-band (right) spectra of the nuclear
  starburst in $\Arp{220}$.
Shaded bands indicate spectral
regions affected by skylines.
\label{fig.Arp220spectra}
}
\end{figure}

\section{The compact, dusty starburst in Arp\,220}

In \figref{fig.Arp220spectra} we present near-IR H- and K-band spectra
of the compact, dusty starburst in the nucleus of the ULIG $\Arp{220}$.
These spectra provide a remarkable contrast with those shown in
\figref{fig.Antspectra}. In the H-band, the only significant emission
line is that of [$\ion{Fe}{ii}$] which probes the supernova remnants
created in the starburst. However, the Brackett series is totally
absent, except for the $\Brg$ line in K-band, which is however dominated
by $\Htwo$ rovibrational lines. Strong photospheric continuum from red 
supergiants is evident in both H- and K-band. The apparent deficiency in
young, hot stars suggested by the faintness of the recombination lines
has led to speculations on a hidden AGN in $\Arp{220}$ which would
provide most of the bolometric luminosity (\cite{Armusetal95}). Alternatively, 
extreme foreground obscuration has been invoked to suppress the
recombination lines (\cite{Sturmetal96}). While extinction certainly plays
a role, the prominence of red supergiant features and of the
[$\ion{Fe}{ii}$] lines, as well as several other arguments
(\cite{VanDerWerf01}) argue against extreme foreground obscuration.

The most satisfying explanation for the faintness of the recombination lines
is Lyman continuum absorption
by dust within the ionized regions (see also Dopita, these proceeedings). 
If most of the ionizing radiation
is absorbed by dust grains rather than hydrogen atoms, a dust-bounded
(rather than hydrogen-bounded) nebula results, and all tracers of
ionized gas (recombination lines, fine-structure lines, free-free
emission) will be suppressed. If the $\HII$ regions in $\Arp{220}$ are
principally dust-bounded, the
observational properties of $\Arp{220}$ can be accounted for, even
with only moderate extinction.
Since the dust also absorbs far-ultraviolet radiation longwards
of the Lyman limit, the formation of photon-dominated regions is
also suppressed, and the thus the same mechanism can account for
the faintness of the $158\mum$ [$\ion{C}{ii}$] line in
$\Arp{220}$ and other ULIGs~(\cite{Fischeretal99}).
 
Is the $\Arp{220}$ starburst dominated by dust-bounded $\HII$ regions?
The \emph{average\/} molecular gas density in the
$\sim10^{10}\Msun$ nuclear
molecular complex in $\Arp{220}$ is
$\nHtwo\sim2\cdot10^4\pun{cm}{-3}$~(\cite{Scovilleetal97}). The strong
emission from high dipole moment molecules such as CS, HCO$^+$ and HCN
argues for even higher densities: $\sim10^{10}\Msun$ of molecular
gas (i.e., \emph{all\/} of the gas in the nuclear complex) has a density
$\nHtwo\sim10^5\pun{cm}{-3}$~(\cite{Solomonetal90}).
At such densities the ionized nebulae created by hot stars are
\emph{compact\/} or \emph{ultracompact\/} $\HII$ regions, where
50 to 99\% of the Lyman continuum is absorbed by
dust~(\cite{WoodChurchwell89}). Observationally, hydrogen-bounded
and dust-bounded $\HII$ regions can be distinguished by
the quantity $R=\qu{L}{FIR}$/$\qu{L}{\mathrm Br\gamma}$: for a wide range of
parameters, $R<3570$ implies that the nebula is hydrogen-bounded, while
$R>35700$ implies that the nebula is dust-bounded~(\cite{Bottorffetal98}).
For $\Arp{220}$, the $\Brg$ luminosity of $1.3\cdot10^6\Lsun$ 
(\cite{VanDerWerf01}), 
implies $R=1.6\cdot10^5$ assuming an obscuring foreground
screen with
$A_V=20\mg$ (a model consistent with all infrared data). 
The star formation takes place in (ultra)compact $\HII$
regions, where
all of the usual tracers of ionized gas (recombination lines,
fine-structure lines, free-free emission) are \emph{quenched}, not
extincted. While this result significantly complicates the interpretation of
diagnostics of massive star formation in ULIGs, it is save to conclude that
the properties of $\Arp{220}$ can be accounted
for by an intense, and significantly (but not extremely) obscured
starburst. There is no reason to invoke the presence of extreme
extinction, a strongly aged starburst, or
an additional power source in $\Arp{220}$.

\bibliographystyle{apalike}
\chapbblname{VanDerWerfP}
\chapbibliography{%
strings,%
Arp220,%
HIIregions,%
IC342,%
M82,%
NGC253,%
NGC1808,%
NGC4038-4039,%
NGC6240,%
NGC7552,%
photoionization,%
starbursts,%
ULIGs%
}

\end{document}